\documentclass[epj,dvips]{webofc}
\usepackage[varg]{txfonts}   
\providecommand{\tabularnewline}{\\}

\newcommand{\sun}{\odot}
\newcommand{\kmps}{\mathrm{km~s^{-1}}}
\newcommand{\Kelvin}{\mathrm{K}}
\newcommand{\Msun}{\mathrm{M_{\sun}}}
\newcommand{\Rsun}{\mathrm{R_{\sun}}}

\woctitle{Physics at the Magnetospheric Boundary}
\begin{document}
\title{Observable Signatures of Classical T Tauri Stars Accreting in
  an Unstable Regime}
%
%

\author{Ryuichi Kurosawa\inst{1}\fnsep\thanks{\email{kurosawa@mpifr-bonn.mpg.de}} \and
        M.~M. Romanova\inst{2}
}

\institute{Max-Planck-Institut f\"{u}r Radioastronomie, Auf dem
  H\"{u}gel, 69, 5312 Bonn, Germany 
\and
  Department of Astronomy, Cornell University, Ithaca, NY 14853-6801,
  USA
}

\abstract{%
We discuss key observational signatures of Classical T Tauri stars
(CTTSs) accreting through Rayleigh-Taylor instability, which occurs at
the interface between an accretion disk and a stellar
magnetosphere. In this study, the results of global 3-D MHD
simulations of accretion flows, in both stable and unstable regimes,
are used to predict the variability of hydrogen emission lines and
light curves associated with those two distinctive accretion flow
patterns.  In the stable regime, a redshifted absorption component
(RAC) periodically appears in some hydrogen lines, but only during a
fraction of a stellar rotation period.  In the unstable regime, the
RAC is present rather persistently during a whole stellar rotation
period, and its strength varies non-periodically.  The latter is caused by
multiple accreting streams, formed randomly due to the instability,
passing across the line of sight to an observer during one stellar
rotation. This results in the quasi-stationarity appearance of the RAC
because at least one of the accretion stream is
almost always in the line of sight to an observer.  In the stable
regime, two stable hot spots produce a smooth and periodic light curve
that shows only one or two peaks per stellar rotation. In the unstable
regime, multiple hot spots formed on the surface of the star, produce
the stochastic light curve with several peaks per rotation period.
}
\maketitle
\section{Introduction}
\label{intro}

Time-scales of the variability exhibited by Classical T Tauri stars
ranges from seconds to decades (e.g., \cite{herbst02},
\cite{rucin08}). The cause of the variability in different
times-scales are likely associated with different physical mechanisms.
Here, we concentrate on the variability that occurs in the time-scale
of a stellar rotation or less. One obvious cause of the variability in this
time-scale is a stellar rotation. For example, the flux modulations
cased by stellar spots, and the obscuration of a stellar surface by a
large-scale warped disk adjacent to the magnetosphere, which corotates
with the star (e.g.~\cite{bouvier03,alencar10,roma13}), are evidently
related with a stellar rotation period. 

An interesting possibility in which matter accretes onto the star via
the magnetic Rayleigh-Taylor (RT) instability has been investigated in
the theoretical studies by e.g., \cite{arons76,spruit93,li:2004}.
More recently, the global three-dimensional (3D) magnetohydrodynamic
(MHD) simulations by e.g.~\cite{roma08,kulk08,kulk09} have shown that
the RT instability induces accretions in multiple (a few to several)
unstable vertically elongated streams or `tongues' which penetrate 
the magnetosphere. They found that the corresponding time-scale of the
variability induced by the instability is typically a few times
smaller than the rotation period of a star.  In this study, we focus
on the variability associated with the unstable accretion caused by
the RT instability, and predict a few key observable signatures of
CTTSs accreting in this regime, by using radiative transfer models
which incorporate the results from the 3D MHD simulations.
The earlier MHD simulations (e.g.~\cite{roma08,kulk08}) have shown that
the unstable tongues are randomly formed, and consequently hot spots on
the stellar surface are also formed irregularly, in time, shapes and
locations. This naturally leads to a formation of an irregular
light-curve. Interestingly, such irregular light-curves are frequently 
observed in CTTSs (e.g.~\cite{herbst94,rucin08,alencar10}). Our aim
here is to find unique signatures of CTTSs accreting in the unstable
regime by means of time-dependent modeling and analysis of emission
line profiles of hydrogen.

The work presented here is based on the studies published in
\cite{kurosawa13}.  In Sec.~\ref{models}, we briefly summarize our
MHD and radiative transfer models. Our main results are presented in
Sec.~\ref{results}, and our main findings are given in
Sec.~\ref{summary}.

\section{Models}
\label{models}

\begin{figure*}
  \begin{center}
    \begin{tabular}{cc}
      \includegraphics[clip,width=0.38\textwidth]{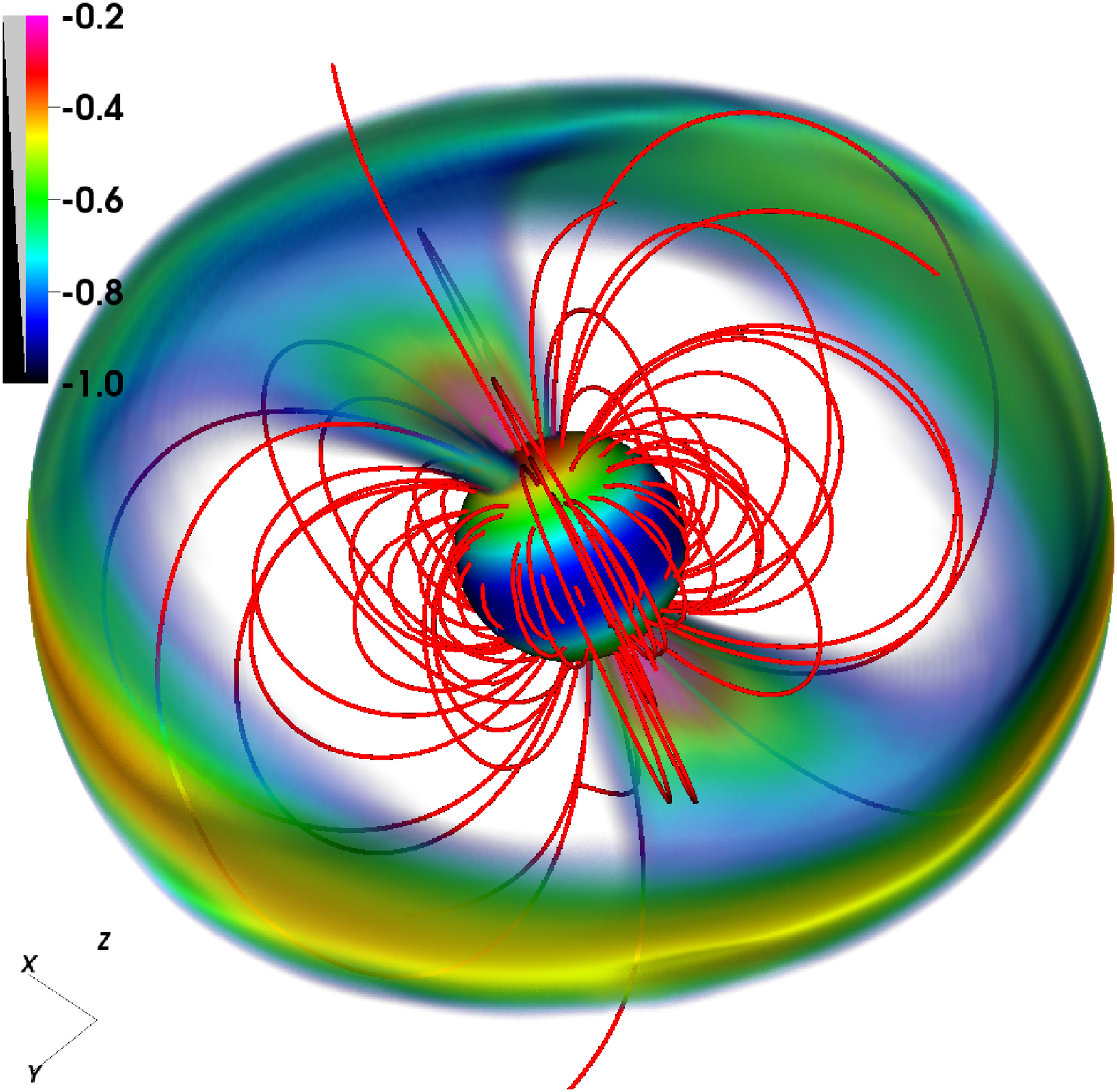} \hspace{1.3cm} &
      \includegraphics[clip,width=0.38\textwidth]{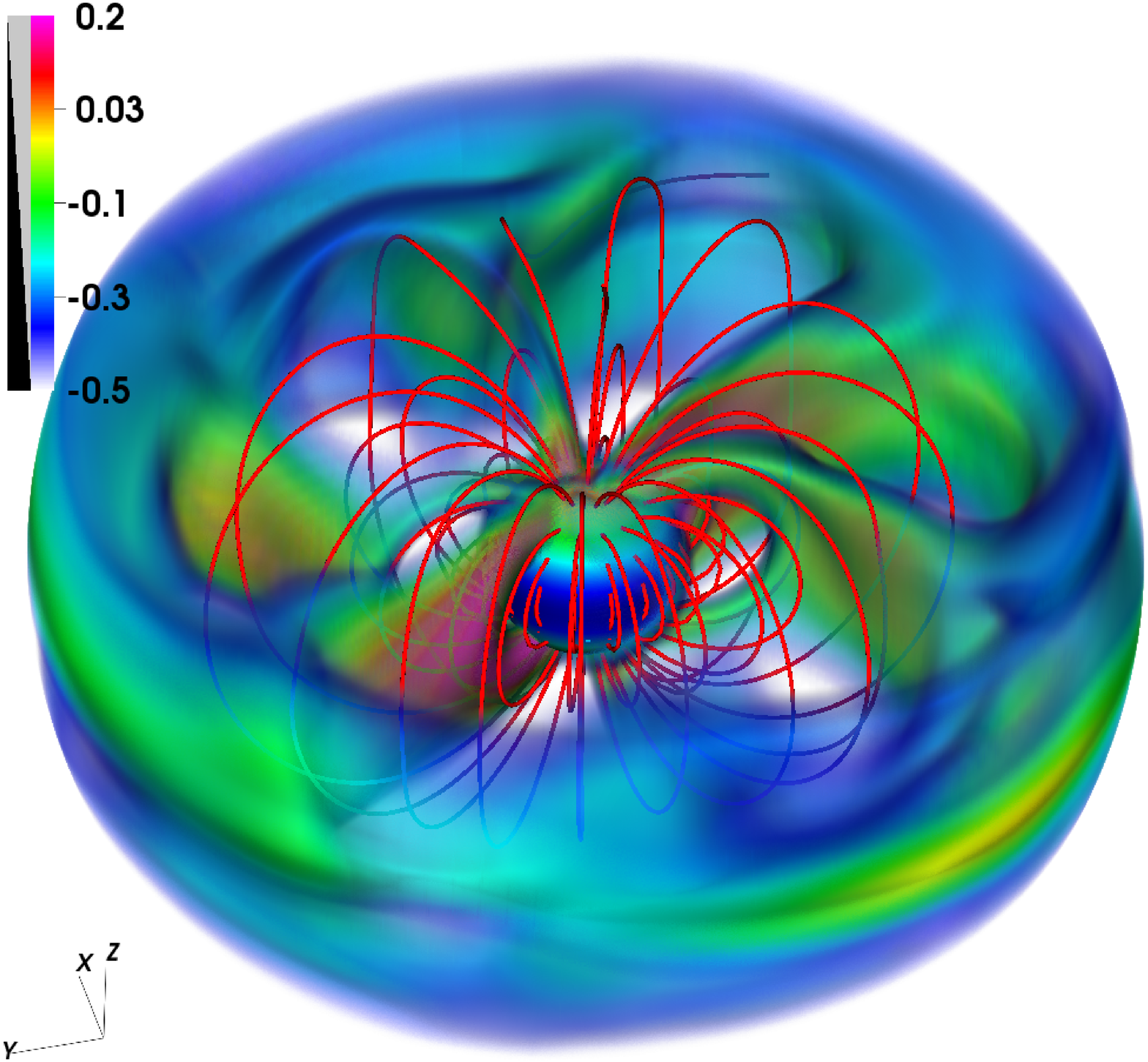}\tabularnewline
    \end{tabular}
  \end{center}
\vspace{-0.5cm}
\caption{Examples of density distributions (volume renderings; in
  logarithmic scales; in arbitrary units) from the 3D MHD simulations
  of the accretion onto CTTSs in a stable (left) and an unstable
  (right) regimes.  Sample magnetic field lines are also shown (red
  lines).
 }
\label{mhd}
\end{figure*}

\begin{table}
\begin{center}
\begin{tabular}{lccccccc}
\hline
         & $M_{*}$ & $R_{*}$ & $B_{\mathrm{eq}}$ & $r_{\mathrm{cor}}$ & $\mathrm{P_{*}}$ & $\Theta$ & $\alpha$\tabularnewline
         & $\left(\Msun\right)$ & $\left(\Rsun\right)$ & $\left(\mathrm{G}\right)$ & $\left(R_{*}\right)$ & $\left(\mathrm{d}\right)$ & $\left(-\right)$ & $\left(-\right)$\tabularnewline
\hline
Stable Case   & $0.8$ & $2$ & $10^{3}$ & $5.1$ & $4.3$ & $30^{\circ}$ & $0.02$\tabularnewline
Unstable Case & $0.8$ & $2$ & $10^{3}$ & $8.6$ & $9.2$ & $5^{\circ}$  & $0.1$\tabularnewline
\hline
\end{tabular}
\caption{Basic model parameters.}
\label{tab:refval}
\end{center}
\end{table}

First, we calculate the matter flows in a global 3D MHD simulation
with frequent writings of data (about 25 time-slices per stellar
rotation and for 3 rotation periods).  The \emph{multiple time-slices}
of the MHD simulation data are then used as the input of a separate
3D radiative transfer code to compute light curves and line profiles which can
be compared with observations.  The multiple time-slices allow us to
follow the time evolution of the line profiles that occurs as a
consequence of the dynamical nature of the accretion flows in the
unstable regime.  In earlier studies, we have used a similar procedure
to calculate line profiles of a star accreting in a stable regime by
using only \emph{a single time-slice} of MHD simulations in which the
gas accretes in two ordered funnel streams \cite{kurosawa08}. More
recently, we have performed a similar modeling for a star with
realistic stellar parameters (i.e.~V2129~Oph), and have found a good
agreement between the model and the time-series observed line profiles
\cite{alencar12}.  

We perform two separate numerical MHD simulations of matter flows with
the parameters shown in Table~\ref{tab:refval}, which are known to
produce a stable and an unstable magnetospheric accretion
(e.g.~\cite{kulk08}). Examples of the magnetospheric accretion flows
in both stable and unstable regimes are shown in Fig.~\ref{mhd}.  The
simulations follow the accretion flows around a rotating star with a
dipole magnetic field with the magnetic axis tilted from the
rotational axis by an angle $\Theta$. The rotational axis of the star
coincides with that of the accretion disk. A viscosity term has been
added to the model with a viscosity coefficient proportional to the
$\alpha$ parameter (e.g.~\cite{shakura:1973}). The detail studies on
the criteria for the magnetic RT instability are presented in e.g.,
\cite{arons76,kaisig92,spruit95}. In general, with our simulation
setup, we find the instability develops more easily when a system has
a relatively smaller tilt angle ($\Theta$), a larger viscosity
($\alpha$) and a longer rotation period ($\mathrm{P}_{*}$) (see
Table~\ref{tab:refval}).

For the calculations of hydrogen emission line profiles from the
matter flow in the MHD simulations, we use the radiative transfer code
\textsc{torus} (e.g.~\cite{harries2000, kurosawa11, kurosawa12}). In
particular, the numerical method used in the current work is
essentially identical to that in \cite{kurosawa11}; hence, for more
comprehensive descriptions of our method, readers are referred to the
earlier papers.  The basic steps for computing the line variability
are as follows: (1)~mapping the MHD simulation data onto the
radiative transfer grid, (2)~source function ($S_{\nu}$) calculations,
and (3)~observed line profile calculations as a function of rotational
phase. In step~(1), we use an adaptive mesh refinement which allows
for an accurate mapping of the original MHD simulation data onto the
radiative transfer grid. The density and velocity values from the MHD
simulations are mapped here. In step~(2), we use a method similar to
that of \cite{klein78} in which the Sobolev approximation
(e.g.~\cite{sobolev1957}) is applied.  The size and shapes
of the hot spots, formed at the stellar surface at the bases of the
accretion streams, are adjusted accordingly with the energy fluxes of
the streams impacting on the stellar surface. The radiation from the
hot spots are included in the level population and the observed flux
calculations. In general, the light curves and line profiles are
greatly influenced by the radiation from hot spots.  We set the gas
temperatures in the accretion funnels for both the stable and unstable
cases to be between $\sim 6000\,\Kelvin$ and $\sim 7500\,\Kelvin$. In
all the variability calculations presented in this work, we adopt the
intermediate inclination angle $i=60^{\circ}$.

\section{Results}
\label{results}

\begin{figure*}
  \begin{center}
    \includegraphics[clip,width=0.84\textwidth]{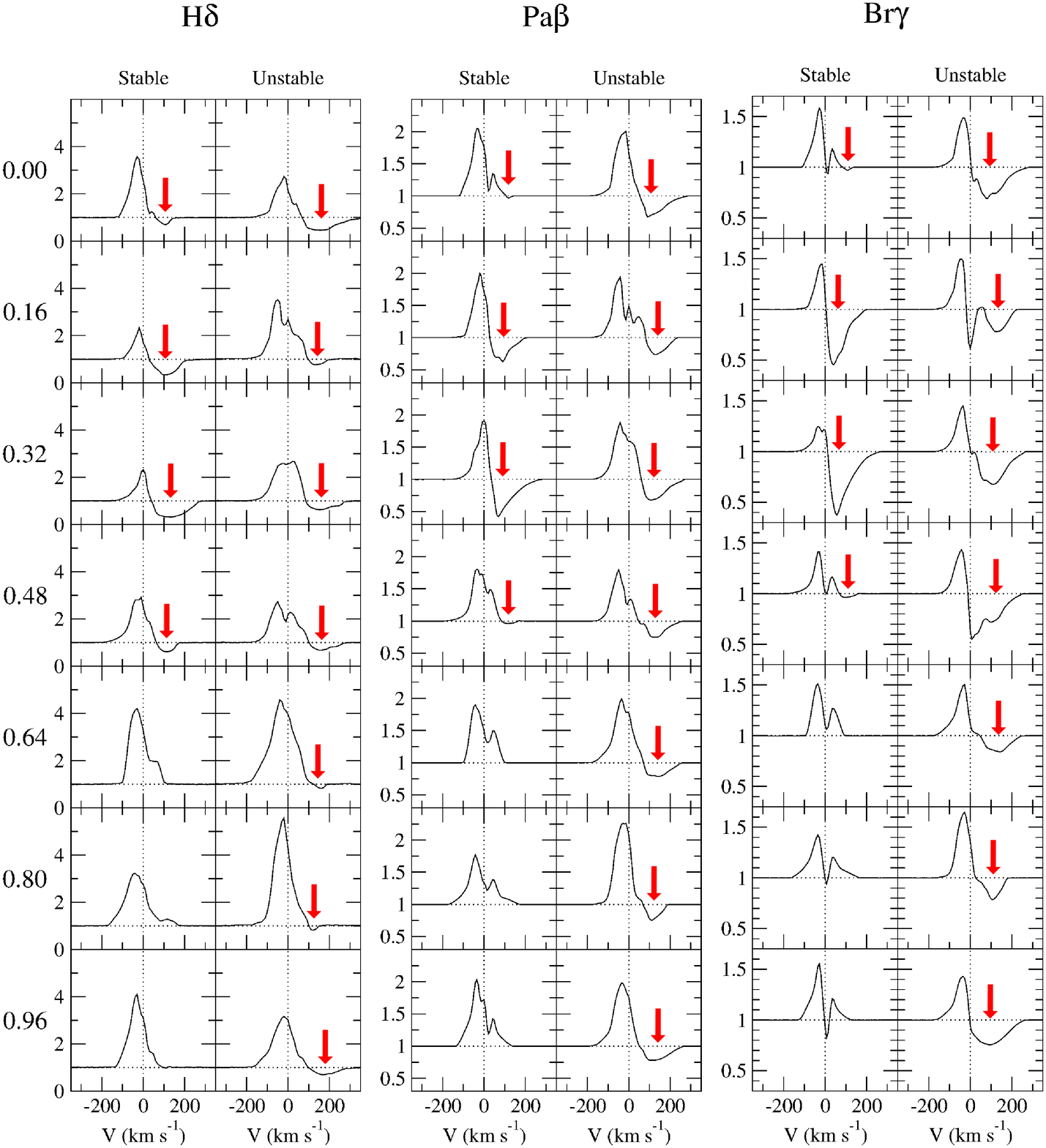} 
  \end{center}
\vspace{-0.5cm}
\caption{The time-series line profile models (solid) of H$\delta$,
  Pa$\beta$ and Br$\gamma$ in the stable regime (left columns) and in
  the unstable regime (right columns) shown for different rotational
  phases (between 0 and 0.96). The rotational phase values are
  indicated in the leftmost column. All the profiles are computed with
  the system inclination angle $i=60^{\circ}$. The red arrows indicate
  the presence and location of the redshifted absorption component.
}
\label{model-profiles}
\end{figure*}

The subsets of the time-series line profile calculations for
H$\delta$, Pa$\beta$ and Br$\gamma$ are summarized in
Fig. \ref{model-profiles}, with the rotational phase step
$\Delta\phi=0.16$, for the first rotation period. See
\cite{kurosawa13} for the models of other hydrogen lines.  The figure
shows that the redshifted absorption components are prominent in
H$\delta$ (in optical), Pa$\beta$ and Br$\gamma$ (in near-infrared)
for both stable and unstable cases. For the stable accretion case, the
redshifted absorption component (RAC) is visible in these lines during
about a half of a rotational phase (between $\phi\sim 0$ and $\sim
0.5$). These phases approximately coincide with the phases when the
upper accretion funnel stream is in front of the star (see
Fig.~\ref{mhd}), i.e.\,when the upper accretion funnel is in the line
of sight of the observer to the stellar surface. A weak double-peak
feature visible in Pa$\beta$ and Br$\gamma$ in the latter half of the
rotational phase ($\phi\sim 0.5$ and $\sim 1.0$) is mainly caused by
the `lack of emission' from the gas moving at zero project velocity
toward an observer. This is due to the geometrical effect (two
distinct accretion streams) and the fact that the emission in
Pa$\beta$ and Br$\gamma$ occurs near the base of accretion stream
where the flow velocity is large ($\sim 200\,\kmps$). At some phases,
the gas at the inner edge of the accretion disk could also intersect
the line of sight between the observer and the system, and causes a
small amount of absorption near the line center since the intervening
gas in the inner most part of the accretion disk is just rotating with
a Keplerian velocity; hence, its projected velocity toward the
observer would be around zero. This can also contribute to the double
peaked appearances of Pa$\beta$ and Br$\gamma$.  For the unstable
case, the RAC in H$\delta$, Pa$\beta$ and Br$\gamma$ is present at
almost all phases. This is due to the geometry of the accretion flows
(see Fig.~\ref{mhd}). The instability causes a few to several
accretion streams to be present in the system at all times.  Hence, at
least one of the accretion streams or tongues is almost always in the
line of sight of the observer to the stellar surface, and producing
the RAC at all the rotational phases.

\begin{figure*}
  \begin{center}
    \begin{tabular}{cc}
      \includegraphics[clip,width=0.45\textwidth]{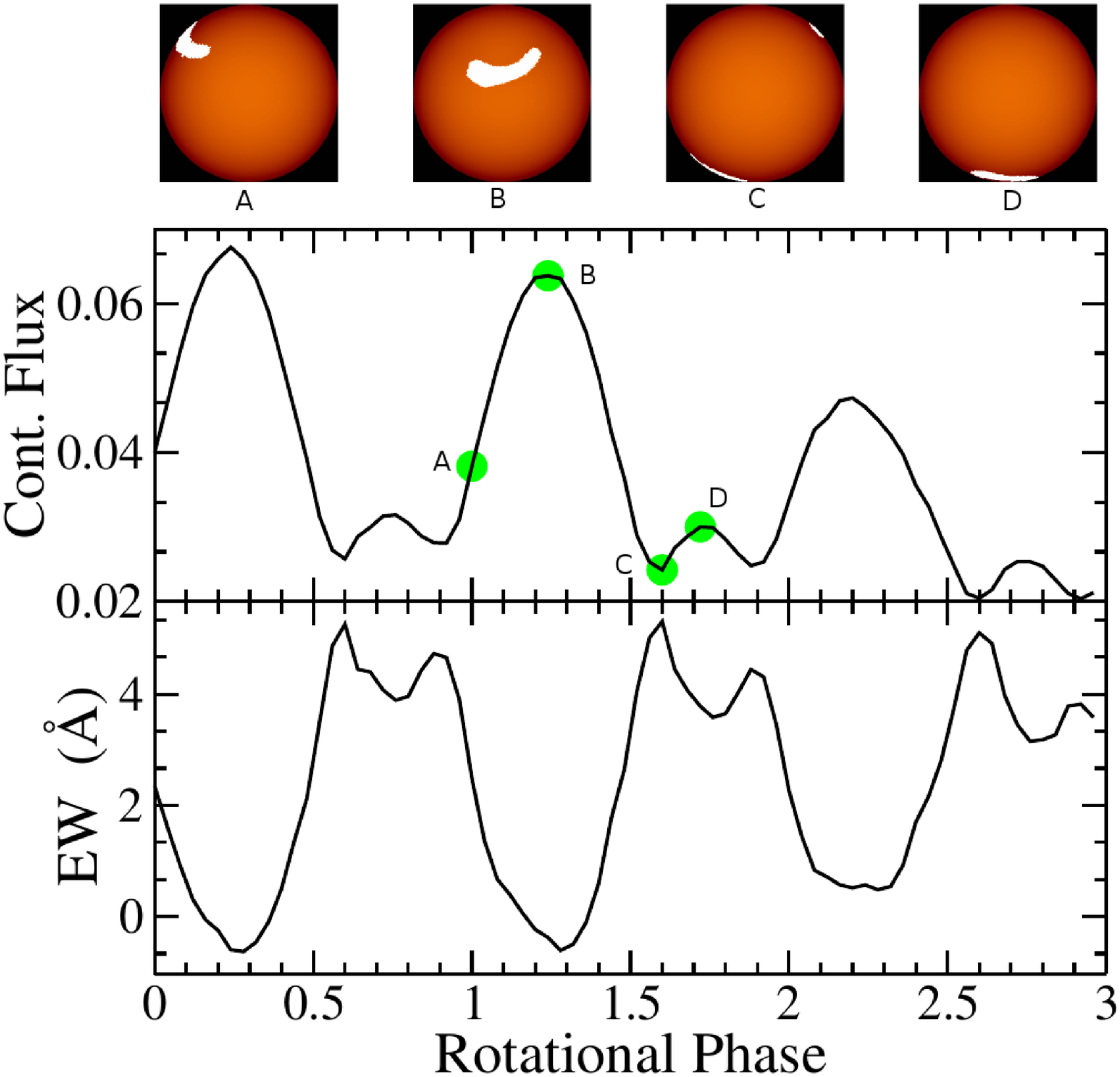} \hspace{0.2cm} &
      \includegraphics[clip,width=0.45\textwidth]{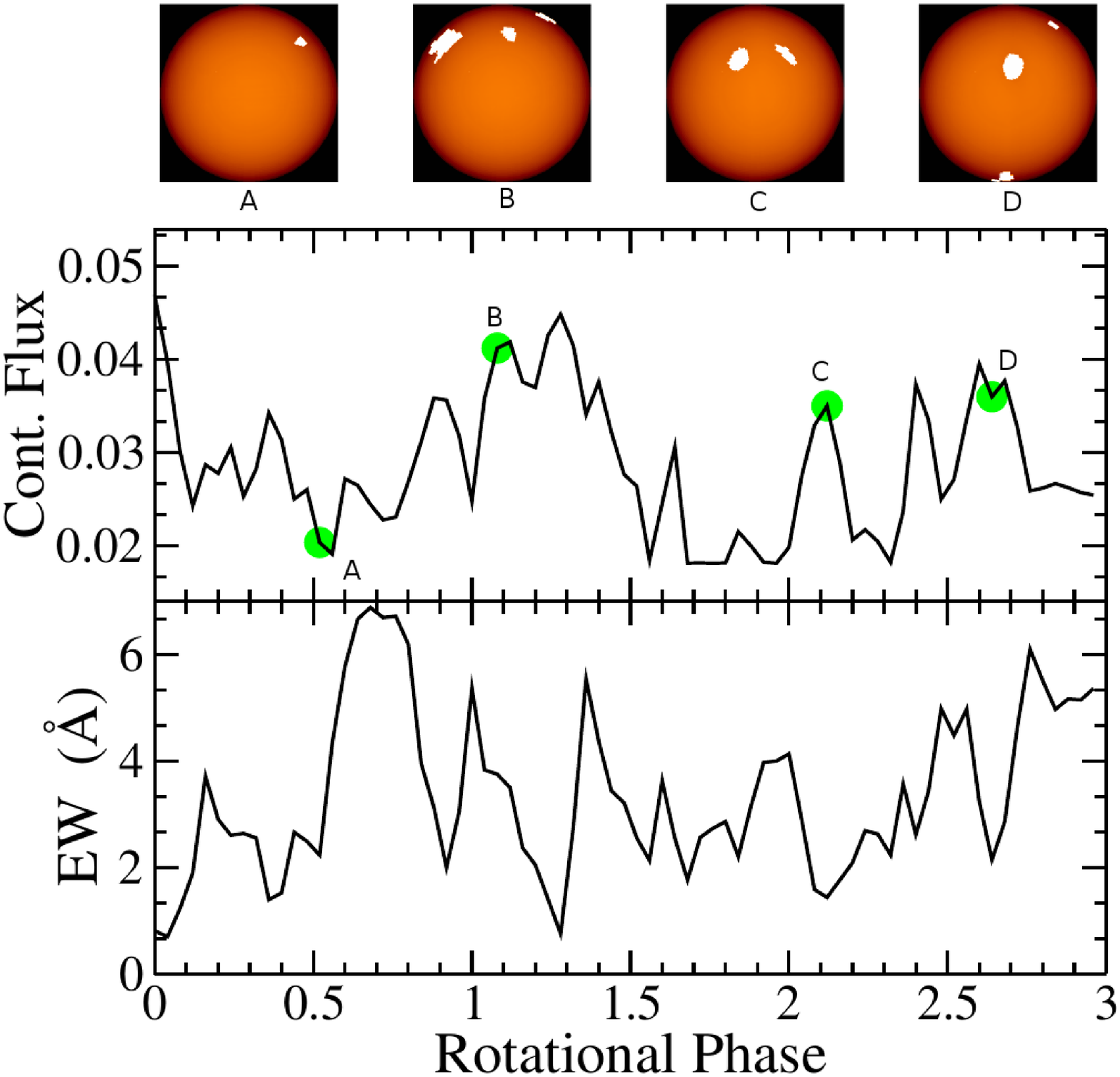}\tabularnewline
    \end{tabular}
  \end{center}
\vspace{-0.5cm}
\caption{The maps of hot spots on the stellar surface at four
  different rotational phases, as seen by an observer (using the
  inclination angle $i=60^{\circ}$).  (top panels), the light-curves
  computed near the wavelength of H$\delta$ (middle panels) and the
  line equivalent widths (EWs) of model H$\delta$ profiles (lower
  panels) are shown for the stable (left panels) and unstable
  (right panels) regimes of accretions.  The hot spots are shown for
  representative moments of time which are marked as green filled
  circles on the light curves.
}
\label{HS-LC-EW}
\end{figure*}

In Fig.~\ref{HS-LC-EW}, we summarize the light curves (monochromatic)
computed at the frequencies around H$\delta$ along with their line EWs
as a function of rotational phase (for three rotational phases) for
both stable and unstable cases. While the light curve is very periodic
for the stable case, it is quite irregular for the unstable
case.  In the stable case, the light curve clearly shows two maxima
per period. These are caused by the rotation of the two hot spots on
the stellar surface created by the two accretion streams (see
Fig.~\ref{mhd}). In an unstable regime, the light curves are
irregular because the accretion streams/tongues form stochastically
and deposit the gas onto the stellar surface in a stochastic
manner. This results in the stochastic formation of hot spots which
produce the stochastic light curve. The line EW variablity curves are
also influenced by the presence of hot spots since they contribute
significantly to the total continuum flux. Since the lines are formed
in the accretion stream which are formed irregularly, their EWs varies
irregularly, as shown in Fig.~\ref{HS-LC-EW}. No clear periodicity is
found in the light curve and the EW curve for the unstable case. 

Interestingly, the regular/periodic and irregular light curves seen in
our models are very similar to some of those found in the observations
by e.g.  \cite{rucin08} and \cite{alencar10}. Note that irregular
light curves are found in about 39~per~cent of the CTTSs sample in
\cite{alencar10}.  No clear periodicity in the line variability is
found in many CTTSs. For example, in the observation of TW~Hya,
\cite{donati11} found that the variability of H$\alpha$ and H$\beta$
are not periodic, and suggested the cause of the variability is
intrinsic (e.g.~changes in the mass-accretion rates) rather than the
stellar rotation.  In the studies of DR~Tau by \cite{alencar01} and
DF~Tau by \cite{johns-krull97}, no clear periodicity was found in
their line variability observations. However, the line variability
seen in these two objects may be also influenced by strong winds
clearly present in some line profiles.

\section{Summary}
\label{summary} 
In this study, we find that the CTTSs accreting in the unstable regime
(due to the magnetic RT instability) would exhibit (1).~irregular
light curves and (2). variable but rather persistent redshifted
absorption component in higher Balmer lines (e.g., H$\gamma$ and
H$\delta$) and in some near-infrared hydrogen lines such as Pa$\beta$
and Br$\gamma$. Although an irregular curve could be also formed by
stochastically occurring flare events, the persistent redshifted
absorption component is not expected to be associated in such events.

\section*{Acknowledgment}
We thank the conference organizers, especially the chair,
Prof. E. Bozzo for the excellent meeting. Resources supporting
this work were provided by the NASA High-End Computing (HEC) Program
through the NASA Advanced Supercomputing (NAS) Division at Ames
Research Center and the NASA Center for Computational Sciences (NCCS) 
at Goddard Space Flight Center.  The research was supported by NASA
grants NNX10AF63G, NNX11AF33G and NSF grant AST-1008636.

%
%
\bibliography{local}


\end{document}